\documentclass[%
 reprint,
 amsmath,amssymb,
 aps,
]{revtex4-2}
\usepackage{braket}
\usepackage{graphicx}
\usepackage{dcolumn}
\usepackage{bm}
\usepackage{hyperref}

\begin{document}
\preprint{APS/123-QED}
\title{\textbf{ Tri-Partite Entanglement In Neutrino Oscillations}}
\author{Abhishek Kumar Jha}\thanks{email: abhiecc.jha@gmail.com}
\author{Supratik Mukherjee}\thanks{email:supratikdgc@gmail.com}
\author{Bindu A. Bambah} \thanks{email: bbambah@gmail.com}
\affiliation{\small{School of Physics, University of Hyderabad, Hyderabad - $500046$, India}}
\date{\today}

\begin{abstract}
We investigate and quantify various measures of bipartite  and tripartite  entanglement in the context of two and three flavor neutrino oscillations. The bipartite entanglement is analogous to the entanglement swapping resulting from a beam splitter in quantum optics. For the three neutrino systems various measures of tripartite entanglement are explored. The significant result is that a monogamy inequality in terms of negativity leads to a residual entanglement, implying true tripartite entanglement in the three neutrino system. This leads us to an analogy of the three neutrino state with a generalized class of W-state in quantum optics.
\end{abstract}

\maketitle

\section{Introduction}
Neutrino Oscillations, the quantum phenomenon in which a neutrino in a given flavor state has a probability of being found in a different flavor state as it progresses in time, are the subject of intense experimental and theoretical activity today \cite{Bilenky:2004,Kayser:2012,Ahmad:2002, Fukuda:2002pe,An:2012eh,Ahn:2006zza,Adamson:2011qu,Abe:2011sj,nova}. Oscillations arise because the neutrino flavor state is  a linear superposition of non-degenerate mass eigenstates of neutrinos. Since quantum entanglement and coherence are two fundamental features arising from the principle of quantum superposition \cite{Bohm}, it is natural to examine quantum entanglement in neutrino systems \cite{Lu-2018, Horodecki:2009zz, Streltsov:2016iow}. Recently, in the context of two flavors, the linear superposition state of a neutrino was examined as a two-qubit system \cite{Blasone:2007vw}. However, not much work has been done for the three flavor system as a three-qubit system, although initial work in this direction was started by Blasone et.al \cite{Blasone:2014jea}. In this paper, we examine, in detail, the entanglement properties of the three-particle superposed flavor-neutrino state. We show that the three neutrino state shows the remarkable property of having a genuine form of three way entanglement akin to the W-state in quantum optics, which can be quantified in terms of appearance and disappearance probabilities.

The coherent time evolution of the neutrinos implies that there is mode entanglement between the mass eigenstates which make up a flavor state \cite{Alok:2014gya}. Due to the “very weakly interacting” character of neutrinos, mode entanglement is the natural form of entanglement to consider for neutrinos. This is often referred to as occupation number entanglement \cite{Blasone:2007vw, Blasone:2014jea, Blasone:2007wp, Blasone:2010ta}. Neutrino entanglement has been studied in both modes, flavor mode as well as in mass mode. Recently, the quantum correlations such as Bell's inequality and Bell-CHSH (Clauser-Horn-Shimoy-Holt) inequality violations, teleportation fidelity and geometric discord for bipartite quantum system have been studied in the context of two flavor neutrino oscillations and related to the neutrino oscillation probabilities \cite{Alok:2014gya}. A temporal analogue of Bell's inequality, the Leggett-Garg inequality, and a quantum information theoretic quantity sensitive to the neutrino mass-hierarchy and similar to the Leggett-Garg inequality have been tested  in the context of three-flavour neutrino oscillations \cite{Formaggio:2016cuh, Naikoo:2017fos, Banerjee:2015mha}. 

 In this paper we show that the behaviour of the two- mode entangled neutrino state resembles entanglement swapping  (the procedure of entangling photons without direct interaction) between two photon states emerging from a Beam Splitter (BS) \cite{Kim:2002, Agarwal:2006, 2005PhRvA..72f4306V, van Enk:2006}. Using this resemblance, we extend the study to three-flavor neutrino oscillations by considering distributed entanglement (i.e, how various measures of entanglement are distributed among the three neutrino states) \cite{Coffman:1999jd}. In quantum optics, this means that if we consider distributed entanglement among pure three qubit states (A, B and C), we can show quantum correlations between A and B, plus between A and C, is either less than or equal to the quantum correlation between A and pair BC (treating it as a single quantum object). These quantum correlations are Tangle, Concurrence and Negativity \cite{Ou_2007}. In order to classify these pure three-qubit states under stochastic local operations and classical communication (SLOCC), we need six classes: (1) The A-B-C class including product states; (2) A-BC, (3) B-AC and (4) C-AB class including bipartite entanglement states; and (5) W and (6) GHZ (Greene-Horn-Zeilinger) classes to study genuine tripartite entanglement in quantum optics \cite{Dur:2000zz}. The W-state is the representative of one of the two non-biseparable classes of three-qubit states (the other being the GHZ-state), which cannot be transformed (not even probabilistically) into each other by local quantum operations. Thus, W and GHZ  represent two very different kinds of tripartite entanglement. Further, the W-state has an interesting property that if one of the three qubits is lost, the state of remaining two-qubit system is still entangled. This robustness of W-type entanglement contrasts strongly with GHZ-state, which is fully separable after the loss of one qubit. Due to this feature, an idea of distributed entanglement, by incorporating a monogamy inequality known as Coffman-Kundu-Wooters (CKW) inequality (using positive partial transposition or reduced density matrix as mathematical tools) of W-class and GHZ-state has been proposed \cite{Coffman:1999jd, Ou_2007}. These two classes of three qubit entangled states (W and GHZ) have given a stimulus to quantum technology by recognizing that these class of states have several applications in quantum information processing - such as quantum teleportation, superdense coding, quantum cryptography etc, in order to build a quantum computer. We show that the three neutrino state, like the W-state, has more robust entanglement. A three-qubit entangled W-state can be produced in a laboratory, using spontaneous parametric down-conversion (SPDC) and two BS \cite{Eibl:2004} (similarly GHZ-state can also be produced via experiments \cite{Pan:2000}). Since we can produce quantum optical states in the lab, the analogy between W-states and the three neutrino state would make it possible to explore the nature of entanglement in neutrino oscillations in the lab.

\section{Entanglement in two-flavor neutrino oscillations}
 The three neutrino flavors, electron neutrino ($\nu_e$), muon neutrino ($\nu_{\mu}$) and tau neutrino ($\nu_{\tau}$) are not neutrino mass eigenstates but a linear superposition of them given by 
 \begin{equation}
\ket{\nu_{\alpha}}=\sum_{j}U_{\alpha j}\ket{\nu_j},
 \end{equation}
where, $\ket{\nu_{\alpha}}$ ($\alpha=e,\mu,\tau$) are the flavor eigenstates, $\ket{\nu_j}$ ($j=1,2,3$) are the mass eigenstates and $U_{\alpha j}$ are the elements of a leptonic mixing matrix called the PMNS (Pontecorvo-Maki-Nakagawa-Sakita) matrix, characterized by three mixing angles ($\theta_{12}$, $\theta_{13}$, $\theta_{23}$) and a charge conjugation and parity (CP) violating phase $\delta_{CP}$\cite{Giganti:2017fhf}. 
${\bf U(\theta_{ij},\delta)}=$

\begin{equation} 
 \begin{pmatrix} c_{12} c_{13} & s_{12} c_{13} & s_{13}e^{-i\delta_{CP}}\\
 -s_{12} c_{23} & c_{12} c_{23} -s_{12} s_{13} & c_{13} s_{23}\\
 -c_{12} s_{13} s_{23}e^{i\delta_{CP}} & s_{23}e^{i\delta_{CP}} &  \\
 s_{13} s_{23}   & -c_{12} s_{23} -s_{12} s_{13} & c_{13} s_{23} \\ 
 -c_{12} s_{13} c_{23}e^{i\delta_{CP}} & c_{23}e^{i\delta_{CP}}  &  
 \end{pmatrix},
 \end{equation}
 where, $c_{ij}= \cos{\theta_{ij}}$ and
 $s_{ij}=\sin{\theta_{ij}}$ ($i,j=1,2,3$). 
 
 In the plane wave picture, the time evolution of the flavor neutrino state is
  \begin{equation}
      \ket{\nu_{\alpha}(t)} =\sum_{j} e^{-iE_j t} U_{\alpha j}\ket{\nu_j}
  \end{equation}
  where, ${U}_{\alpha\beta}(t)\equiv\sum_j U_{\alpha j} e^{-iE_j t} U^*_{\beta j}$ and $E_j$ is the energy associated with the mass eigenstate $\ket{\nu_j}$. Consequently, the probability for detecting another flavor neutrino $\beta=(e,\mu,\tau)$ from an initial $\alpha$ neutrino is
   \begin{eqnarray}
&& P_{\alpha\beta} = \delta_{\alpha\beta} - 4 \sum_{j>k} Re(U^*_{\alpha j} U_{\beta j} U_{\alpha k} U^*_{\beta k}) \sin^2 \left(\Delta m^2_{jk} \frac{Lc^3}{4\hslash E} \right) \nonumber \\
&& \hspace{1em}   + 2\sum_{j>k} Im(U^*_{ej} U_{\beta j} U_{e k} U^*_{\beta k} ) \sin^2 \left(\Delta m^2_{jk} \frac{Lc^3}{2\hslash E} \right)
   \end{eqnarray}
    where $\Delta m^2_{jk}\equiv m^2_j -m^2_k$, E is the energy of the neutrino which is different for different neutrino experiments and $L$ is the distance traveled by the neutrino from source to detector \cite{Gonzalez-Garcia:2014bfa}. 
  
  Thus, using Eq.(3), the evolved neutrino flavor state in a coherent superposition of flavor basis can be written as,
  \begin{equation}
 \ket{\nu_\alpha (t)}=\tilde{U}_{\alpha e}(t)\ket{\nu_e}+\tilde{U}_{\alpha \mu}(t)\ket{\nu_\mu} + \tilde{U}_{\alpha \tau}(t)\ket{\nu_\tau} 
 \end{equation}
 
 First, we characterize two qubit entanglement for two-flavor mixing which are relevant, as a first approximation, to three cases of neutrino experiments. $\nu_\mu\leftrightarrow\nu_{\tau}$ transitions are relevant for atmospheric neutrinos, $\nu_e\leftrightarrow\nu_{\mu}$ at reactor experiments
  and $\nu_{\mu}\leftrightarrow\nu_{e}$ at accelerator experiments \cite{Adamson:2011qu, An:2012eh, nova, Abe:2017uxa}. For electron and muon neutrino entanglement, we identify 2 qubit states with the flavor state at time t=0 by using the occupation number states as \cite{Blasone:2007vw} 
 \begin{eqnarray}
  \ket{\nu_{e}}=\ket{1}_e \otimes \ket{0}_\mu \equiv \ket{10}_e , \nonumber \\
 \ket{\nu_{\mu}}=\ket{0}_e \otimes \ket{1}_\mu\equiv \ket{01}_\mu . \nonumber 
  \end{eqnarray}
  For two neutrino mixing the SU(2) rotation matrix
  \begin{eqnarray}R(\theta)=\begin{pmatrix}
 \cos\theta & \sin\theta\\
 -\sin\theta & \cos\theta\\
 \end{pmatrix},\end{eqnarray} \\
 can be identified with the mixing matrix $U(\theta)$.

 The time evolution of an initial electron-flavor neutrino state in two mode (flavor) system can be obtained from Eq.(5) as,
  \begin{equation}
   \ket{\nu_e(t)}=\tilde{U}_{ee}(t)\ket{10}_e+\tilde{U}_{e\mu}(t)\ket{01}_\mu .
  \end{equation}
  
 The probability of generating and detecting electron-neutrino flavor state as a disappearance probability $P_d=\vert{\tilde{U}_{ee}(t)}\vert^2$ and, the probability of generating electron-neutrino flavor state and detecting muon-neutrino flavor state as an appearance probability $P_a=\vert{\tilde{U}_{e\mu}(t)}\vert^2$ are,
 
 \begin{eqnarray}
& {P_{d}}= {\cos^4 \theta}+ {\sin^4 \theta }+ 2\,{\sin^2 \theta}\, {\cos^2 \theta} \cos\left(\dfrac{\Delta m^2 t}{2E}\right) &\\ 
& \mbox{and} \hspace{1em} {P_{a}}= 4\,{\sin^2 \theta}\, {\cos^2 \theta} \sin^2\left(\dfrac{\Delta m^2 t}{4E}\right). &
\end{eqnarray}

where $\theta$ is a generic two flavor mixing angle and $\Delta m^2$ is the corresponding mass-square difference. The corresponding density matrix $\rho^{e\mu}(t)$ is given by
 $\rho^{e\mu} (t) = \ket{\nu_e(t)} \bra{\nu_e (t)}$ such that,
 \begin{equation}
 \rho^{e\mu} (t)=\begin{pmatrix}
 0 & 0 & 0 & 0\\
 0 & \vert{\tilde{U}_{ee}(t)}\vert^2 & \tilde{U}_{ee}(t) \tilde{U}_{e\mu}^*(t) & 0\\
 0 &  \tilde{U}_{e\mu}(t)\tilde{U}_{ee}^*(t) & \vert{\tilde{U}_{e\mu}(t)}\vert^2 & 0 \\
 0 & 0 & 0 & 0\\
  \end{pmatrix}
 \end{equation}

A good optical analogy to the phenomenon of neutrino  oscillation is the following situation. In quantum optics, the action of a quantum mechanical BS (Beam splitter) interferometer is given by the $SU(2)$ matrix $R(\theta)$, which performs exactly the same transformation on photons as the neutrino mixing matrix does. Thus, the entanglement in a two flavour neutrino mixing is akin to entanglement via mode swapping that takes place due to a BS \cite{Agarwal:2006}.  

 Let $\rho^{e\mu}(t)$ be a density operator of an initial electron-neutrino flavor state Eq.(7), where $e$ and $\mu$ are electron flavor mode and muon flavor mode respectively, in two-qubit mode (flavor) basis (i.e, $\ket{10}_e$ and $\ket{01}_\mu$ ). A matrix element of the density operator is of the form  $\rho^{e\mu}_{pq,rs} (t) = \bra{p}\bra{q}\rho^{e\mu}(t)\ket{r}\ket{s}$. The partial transpositions of operator $\rho^{e\mu}(t)$ in flavor modes $e$ and $\mu$ are defined as
$\rho^{T_e}_{pq,rs}(t) =\rho^{e\mu}_{rq,ps}(t)$ and
$\rho^{T_\mu}_{pq,rs}(t) =\rho^{e\mu}_{ps,rq}(t)$. The Peres-Horodecki criterion, which is also called as Positive Partial Transpose (PPT) criterion, is a sufficient condition for separability in bipartite quantum system, where the composite state $\rho^{e\mu}(t)$ is separable if and only if $\rho^{T_e}(t)$ or $\rho^{T_\mu}(t)$ is a positive operator, with all positive eigenvalues, otherwise the composite state $\rho^{e\mu}(t)$ is an entangled state \cite{DARIUSZ:2012}. The partial transpose in muon-flavor mode from Eq.(10) is
$\rho^{T_\mu}(t) =$
\begin{equation}
\begin{pmatrix}
 0 & 0 & 0 & \tilde{U}_{ee}(t)\tilde{U}_{e\mu}^*(t)\\
 0 & \vert{\tilde{U}_{ee}}(t)\vert^2 & 0 & 0\\
 0 & 0 & \vert{\tilde{U_{e\mu}}}(t)\vert^2 & 0 \\
 \tilde{U}_{e\mu}(t)\tilde{U}_{ee}^*(t) & 0 & 0 & 0\\
 \end{pmatrix}
\end{equation}
In terms of probabilities, the eigenvalues $\lambda_i$ of $\rho^{T_\mu}(t)$ are $\lambda_1 = P_{d}$, $\lambda_2 =P_{a}$, 
$\lambda_3 =\sqrt{P_{d}P_{a}}$, $\lambda_4 =-\sqrt{P_{d}P_{a}}$. Thus $\lambda_4$ is not positive which means $\rho^{T_\mu}(t)$ is not a positive operator and therefore the $e$ $\mu$ neutrino state is entangled. Negativity is a quantity which measures by how much $\rho^{T_\mu}(t)$ fails to be positive definite \cite{Zyczkowski:1998yd, Vidal:2002zz, Vedral:2005, Malvimat:2018izs}. The condition Negativity $N_{e\mu}>0$ is the necessary and sufficient inseparable condition for the bipartite quantum system to be entangled  and is defined as \begin{equation}
N_{e\mu}=N(\rho^{e\mu} (t)) = \frac{\vert\vert {\rho^{T_\mu}}(t)\vert\vert -1}{2} ,
\end{equation}
where the trace norm 
\begin{equation}
{\vert\vert {\rho^{T_\mu}(t)}\vert\vert} =Tr\sqrt{\rho^{T_\mu}(t){\rho^{T_\mu}}^\dagger (t)} = 1+2{\vert \sum_i {\lambda_i}\vert},
 \end{equation}
and $\lambda_i <0$ are the negative eigenvalues of partial transposition $\rho^{T_\mu}(t)$ \cite{Horodecki:1998kf}. 
For the two flavor neutrino oscillations, \begin{equation}
\vert\vert {\rho^{T_\mu}(t)}\vert\vert 
=1+2{\sqrt{P_{d}P_{a}}}.
\end{equation}
Thus, the negativity is 
$N_{e\mu}=2{\sqrt{P_{a}P_{d}}}$ which is always greater than 0, so $e$-$\mu$ neutrino system is maximally entangled \cite{Ou_2007}.

Non-locality measures like Concurrence and Tangle are strong aspects of quantum correlations \cite{osterloh:2002,Sarandy:2004}. A general  bipartite state $\psi$ of a two qubit system can be written as $\ket{\psi}=A{\ket{10}}+B\ket{01}$, where $\vert A\vert^2 +\vert B\vert^2 =1$. 

\begin{figure}[ht!] 
        \centering \includegraphics[width=1.0\columnwidth]{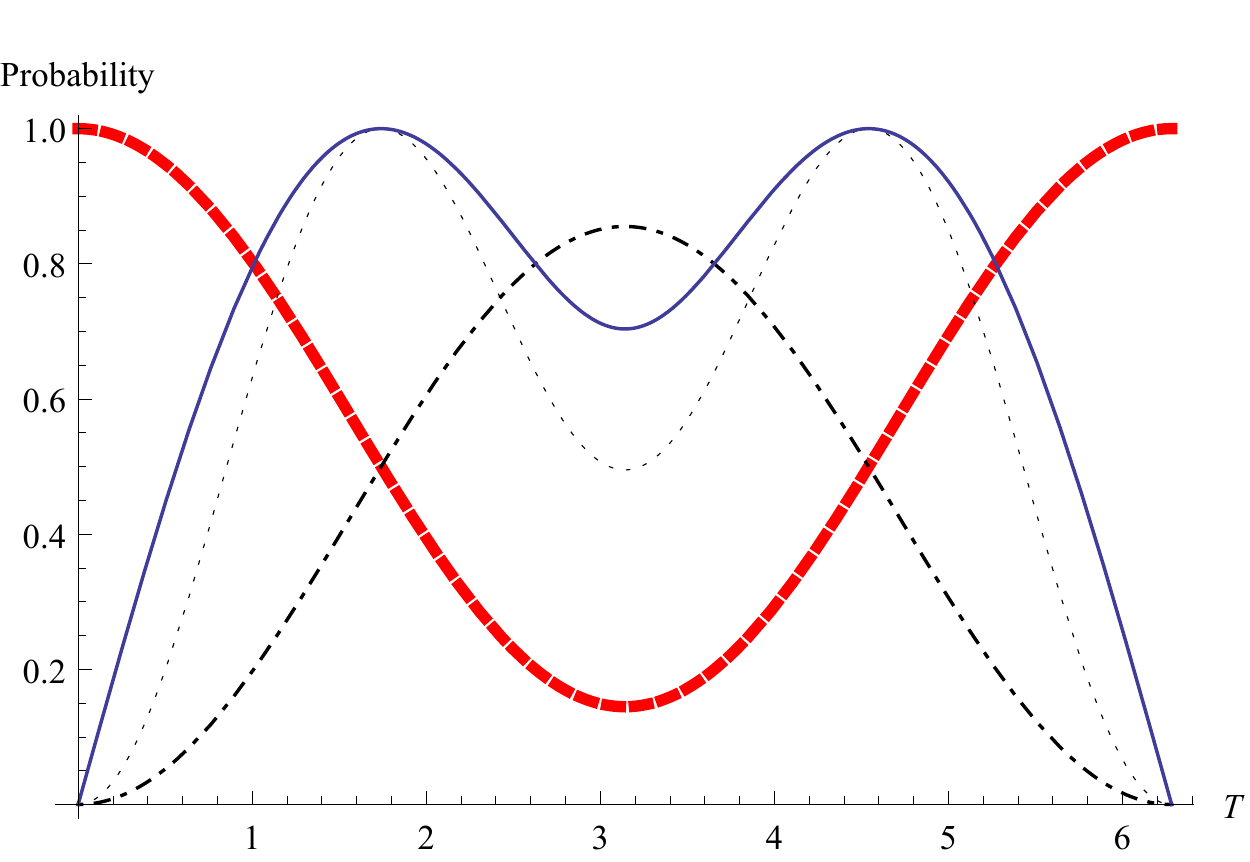}    
        \caption{(Color online) Measures of bipartite quantum correlations Tangle ($\tau_{e\mu}$) = linear entropy ($S_{e\mu}$) (dotted line), Negativity ($N_{e\mu}$) = Concurrence ($C_{e\mu}$) (full line) vs scaled time $T\equiv(\frac{\Delta m^2 t}{2E})$ for an initial electron flavor neutrino state. In order to see the dependence on oscillation probabilities, the transition probabilities $P_d$ (dashed line) and $P_a$ (dotted dash line) are also plotted. The mixing angle $\theta$ is fixed at the experimental value $\sin^2\theta=0.310$ 
 \cite{Esteban:2018azc}}
\label{fig:1}
\end{figure}

The \textquotedblleft {spin-flipped} \textquotedblright    density matrix $\tilde{\rho}^{AB}(t)$ of the state $\ket{\psi}$ is $\tilde{\rho}^{AB}(t)=(\sigma_y \otimes \sigma_y)\rho^{*AB}(t)(\sigma_y \otimes \sigma_y)$, where $\rho^{*AB}(t)$ denotes the complex conjugation in the standard basis ($\ket{00}$,$\ket{01}$,$\ket{10}$,$\ket{11}$) and  $\sigma_x$ and $\sigma_y$ are Pauli matrices. As both ${\rho^{AB}(t)}$ and $\tilde{\rho}^{AB}(t)$ are positive operators, it follows that the product ${\rho^{AB}(t)\tilde{\rho}^{AB}(t)}$, though non-hermitian also has only real and non-negative eigenvalues. Let the square roots of these eigenvalues, in the decreasing order be $\lambda_1$,$\lambda_2$,$\lambda_3$,$\lambda_4$, then the Tangle of the density matrix $\rho^{AB}(t)$ is defined as \cite{Coffman:1999jd}:
 \begin{equation}
 \tau_{AB}=[max(\lambda_1 -\lambda_2 -\lambda_3-\lambda_4 ,0)]^2
 \end{equation}
 
 We identify A and B with the  $e$ and $\mu$ flavor modes respectively. Using Eq.(10), one can show that the tangle is  $\tau_{e\mu} =4det[{\rho^{e}(t)}]$, where $\rho^e (t)$ is the density matrix associated to the reduced state after tracing over muon flavor mode i.e, $\rho^e(t) =Tr_\mu ({\rho^{e\mu}(t)})= \begin{pmatrix}
 \vert{\tilde{U}_{ee}(t)}\vert^2 & 0\\
 0 & \vert{\tilde{U}_{e\mu}(t)}\vert^2\\
 \end{pmatrix}
 $. Therefore, the tangle ($\tau_{e\mu}$) for two flavor neutrino oscillations is:
\textbf{\begin{equation}
\tau_{e\mu}= 4\vert{\tilde{U}_{e\mu}}(t)\vert^2 \vert{\tilde{U_{ee}}}(t)\vert^2 = 
4{P_{a}P_{d}}.
\end{equation}}

 Similarly, Concurrence is a measure of entanglement defined as \cite{Wootters:1997id} :
\begin{equation}
 C_{AB} =[max(\lambda_1 -\lambda_2 -\lambda_3 -\lambda_4 ,0)],
 \end{equation}
which for the  electron-neutrino flavor system is \cite{Klyachko:2006}:
\textbf{\begin{equation}
C_{e\mu} = \sqrt{\tau_{AB}} =2\sqrt{P_{a}P_{d}}.
\end{equation}}
 In the ultra-relativistic approximation, Fig1. shows all measures of bipartite quantum correlations, $\tau_{e\mu}$ (dotted line) and $N_{e\mu}$ = $C_{e\mu}$ (full line), with transition probabilities $P_d$ (dashed line) and $P_a$ (dotted dash line), of an initial electron-neutrino flavor state as a function of scaled time $T\equiv\frac{\Delta m^2 t}{2E}$. The mixing angle $\theta$ and the squared mass differences ($\Delta m^2$) are fixed at the most recent experimental values reported in Ref.\cite{Esteban:2018azc}. At T=0, all measures of entanglement are zero, i.e, $N_{e\mu}$, $\tau_{e\mu}$, and $C_{e\mu}$ corresponds to an unentangled state and the two flavor modes are not mixed. For $T>0$, initial electron-neutrino flavor state exhibits oscillatory behaviour. When transition probabilities is maximum  $P_a=P_d=0.5$, all measure of entanglement tends to 1 i.e, $N_{e\mu}=\tau_{e\mu} =C_{e\mu}=1$, which corresponds to maximally entangled state.
In two flavor neutrino oscillation, among entanglement monotones, linear entropy is linked to the variances associated with the average neutrino number \cite{Blasone:2007vw}. The linear entropy is a lower approximation to the von Neumann entropy and for electron-neutrino flavor state, $S_{e\mu}=4 {P_{a} P_{d}}=\tau_{e\mu}$.  

 We see that all measures of entanglement in bipartite system- Negativity, Concurrence and Tangle are directly proportional to the product of appearance and disappearance probabilities and coincide with linear entropy such that
\textbf{\begin{equation}
N^2 _{e\mu} =C^2 _{e\mu}=\tau_{e\mu}=S_{e\mu}=4P_a P_d.
\end{equation}}
This means that the electron-neutrino flavor state is a pure state and these quantum correlations have a  direct experimental connection with physical quantities in neutrino oscillations \cite{Klyachko:2006}.

At this juncture, we are in a position to compare our single particle neutrino state with a single photon system, where the quantum uncertainty on \textquotedblleft which path\textquotedblright of the photon at the output of an unbalanced Beam Splitter (BS) is replaced by the uncertainty on \textquotedblleft which flavor\textquotedblright of the produced neutrino is measured \cite{Blasone:2007vw}. The coefficients $\tilde{U}_{\alpha e}(t)$ and $\tilde{U}_{\alpha\mu}(t)$ play the role of transmissivity (T) and the reflectivity (R) of the BS, respectively and $BS=R(\theta)\equiv U(\theta)$ (i.e, here BS is identifed as a beam splitter transformation matrix $U(\theta)$), in two-flavor neutrino oscillations. Let us consider the simplest case Eq.(7), when time evolution flavor electron neutrino $\ket{\nu(t)}_e$ enters from the port 1, and no neutrino enters from the port 2, into the BS. The single particle neutrino state take two paths - it either gets transmitted ($T\equiv \tilde{U}_{ee}(t)$) or is reflected ($R\equiv \tilde{U}_{e\mu}(t)$). Thus, the state produced by the $\ket{\nu(t)}_e$ has the form of two mode entangled state ($\ket{10}_e$ and $\ket{01}_{\mu}$); more precisely it is a flavor-entangled state like the Bell's state/two qubit state in quantum optics.

\section{Entanglement in three-flavor neutrino oscillations}
  In the three generation framework of neutrino oscillation system we identify neutrino modes in the occupation number basis as: 
  \begin{eqnarray}
  \ket{\nu_{e}}&=&\ket{1}_e \otimes \ket{0}_\mu \otimes \ket{0}_\tau \equiv \ket{100}_e, \nonumber \\\ket{\nu_{\mu}}&=&\ket{0}_e \otimes \ket{1}_\mu  \otimes \ket{0}_\tau \equiv \ket{010}_\mu,\nonumber\\
 \ket{\nu_{\tau}}&=&\ket{0}_e \otimes \ket{0}_\mu  \otimes \ket{1}_\tau \equiv \ket{001}_\tau.
 \end{eqnarray} 
 In this section, we will explore various quantum correlations like Tangle, Concurrence, and Negativity in the three neutrino system. We study two types of entanglement in the tripartite quantum system. Firstly, we look at pairwise entanglement, treating one flavor mode as one object (e.g $e$) and the other two as single object (e.g $\mu\tau$), and the other two permutations of this system. This is a type of bipartite entanglement in a three flavor (tripartite) system which can  quantified by bipartite measures like the concurrence, tangle and negativity defined earlier. Later, we will consider genuine tripartite entanglement, for which a measure called residual entanglement in terms of tangle, concurrence, and negativity is constructed separately. By calculating the genuine tripartite entanglement, we can check if the neutrino state is a generalized case of the two types of tripartite states in quantum optics, the W-state or the GHZ-state. Only the generalized W-state has a residual entanglement called three-$\pi$ (which will be defined later), the GHZ-state has a zero residual three-$\pi$ i.e, the GHZ-state does not have any physical significance in neutrino oscillations. So, the existence of a non-zero residual entanglement three-$\pi$ will put neutrino states in the same class as W-states.
 
 In a reactor type neutrino experiment where an electron neutrino produced at the source can oscillate into other flavors, using Eq.(5) and Eq.(20), the time evolution of electron-neutrino state, in the occupation number basis can be written as \cite{Blasone:2014jea}
  \begin{equation}
 \ket{\nu_{e}(t)}=\tilde{U}_{ee}(t)\ket{100}_e+\tilde{U}_{e\mu }(t)\ket{010}_\mu+\tilde{U}_{e\tau}\ket{001}_\tau
  \end{equation}  with normalization condition  $\vert{\tilde{U}_{ee}(t)}\vert^2+\vert{\tilde{U}_{e\mu }(t)}\vert^2+\vert{\tilde{U}_{e\tau}(t)}\vert^2=1$,
 where $e$, $\mu$ and $\tau$ are a three modes (flavor) neutrino state $\ket{100}_e$, $\ket{010}_\mu$, and $\ket{001}_\tau$ respectively, in three-qubit system.
 The corresponding density matrix in the standard basis $\ket{ijk}$, where each index takes the values 0 and 1 is given by $\rho^{e\mu\tau}(t)=$
 \begin{eqnarray}
 \begin{pmatrix}
 0 & 0 & 0 & 0 & 0 & 0 & 0 & 0\\
 0 & 0 & 0 & 0 & 0 & 0 & 0 & 0\\
 0 & 0 & 0 & 0 & 0 & 0 & 0 & 0\\   
 0 & 0 & 0 & \vert{\tilde{U}_{ee}(t)}\vert^2 & 0 & \tilde{U}_{ee}(t) \tilde{U}_{e\mu}^*(t) & \tilde{U}_{ee}(t) \tilde{U}_{e\tau}^*(t) & 0 \\
  0 & 0 & 0 & 0 & 0 & 0 & 0 & 0\\
  0 & 0 & 0 & \tilde{U}_{e\mu}(t) \tilde{U}_{ee}^*(t) & 0 & \vert{\tilde{U}_{e\mu} (t)}\vert^2 & \tilde{U}_{e\mu}(t)\tilde{U}_{e\tau}^*(t) & 0 \nonumber \\ 
  0 & 0 & 0 & \tilde{U}_{e\tau}(t) \tilde{U}_{ee}^*(t) & 0 & \tilde{U}_{e\tau}(t) \tilde{U}_{e\mu}^* (t) & \vert{\tilde{U}_{e\tau} (t)}\vert^2 & 0\\
  0 & 0 & 0 & 0 & 0 & 0 & 0 & 0
 \end{pmatrix}. 
 \end{eqnarray}
 
 In pairwise entangled tripartite quantum system, the probability of the three flavour state to be an $e$ neutrino mode is $P_{d} =\vert{\tilde{U}_{ee}(t)}\vert^2$ and to be in the $\mu\tau$ mode (treating as a single quantum object) is $P_{a}=\vert{\tilde{U}_{e\mu} (t)}\vert^2 + \vert{\tilde{U}_{e\tau} (t)}\vert^2$. 
Under the action of the partial transposition operator on the density matrix $\rho^{e\mu\tau}(t)$, the matrix elements change under the rule $\ket{ijk}\bra{i'j'k'}$ $\longrightarrow$ $\ket{i'jk}\bra{ij'k'}$.
The eigenvalues of $\rho^{T_e}(t)$ are,
\begin{eqnarray}
 \lambda_1&=&\lambda_2=\lambda_3=\lambda_4=0,\nonumber \\\lambda_5 &=& P_{d}\\\lambda_6 &=&P_{a} \nonumber \\ \lambda_7&=&\sqrt{P_{a} P_{d}} \nonumber \\\lambda_8 &=&-\sqrt{P_{a} P_{d}}. \nonumber  \end{eqnarray} Thus $\lambda_8 $ is not positive which means $\rho^{T_e}(t)$ is not positive operator and therefore $\rho^{e\mu\tau}(t)$ is entangled with reference to the PPT criterion. 
 
 Consequently, finding $\vert\vert {\rho^{T_e}(t)}\vert\vert=1+2\sqrt{P_{a}P_{d}}$, the negativity is given  $N_{e(\mu\tau)} =2\sqrt{P_{a}P_{d}}$ and is positive, fulfilling the criterion of maximal entanglement.\\
 The reduced density matrix $\rho^e(t)$ after tracing one mode (flavor) is \begin{eqnarray}
 \rho^{e}(t)&=&Tr_{\mu\tau}(\rho^{e\mu\tau}(t)) \nonumber \\
 &=&\begin{pmatrix}
\vert{\tilde{U}_{ee}(t)}\vert^2 & 0\\
0 & \vert{\tilde{U}_{e\mu} (t)}\vert^2 + \vert{\tilde{U}_{e\tau} (t)}\vert^2
\end{pmatrix}.\end{eqnarray} 
The tangle, $\tau_{e(\mu\tau)}=2[1-Tr{(\rho^{e}(t))^2}]=4P_{a}P_{d}$ and the concurrence, $C_{e(\mu\tau)}=2\sqrt{P_{a}P_{d}}$. When neutrino oscillates in between different modes (flavor), the linear entropy of the reduced state is $S_{e(\mu\tau)}=4P_{a}P_{d}$. Hence all measures of quantum correlations of bipartite states of three qubit mode (flavor) entangled single particle neutrino state $\ket{\nu_e (t)}$ are satisfied. They are related by 
\begin{equation}
   N^{2} _{e(\mu\tau)}=C^{2} _{e(\mu\tau)} =\tau_{e(\mu\tau)}=S_{e(\mu\tau)}=4P_a P_d.  
\end{equation}

This analysis shows that the entanglement quantified by the tangle, concurrence and negativity between flavor modes $e$ and $\mu$, between $e$ and $\tau$, and between $e$ and single object $\mu\tau$ for the electron-neutrino flavor state has pairwise bipartite entanglement .

However, in order to understand a genuine tripartite entanglement, the neutrino state should be neither fully separable nor biseparable.
The following criteria have to be met for genuine bipartite entanglement 
\begin{itemize}
\item The quantum correlations in electron-neutrino flavor state Eq.(21) have to satisfy the CKW inequality, which is a monogamy inequality for concurrences:\\$C^{2} _{e\mu}+C^{2} _{e\tau}\leq C^{2} _{e(\mu\tau)}$ 
\item The monogamy inequality for tangles\\
$\tau_{e\mu}+\tau_{e\tau}\leq\tau_{e(\mu\tau)}$,
 \item The monogamy inequality for negativity \\ $N^{2} _{e\mu}+N^{2} _{e\tau}\leq N^{2} _{e(\mu\tau)}$.
 \end{itemize}
 
We can also define three quantities that quantify three particle entanglement called the residue of tangle, concurrence and negativity $\tau_{e\mu\tau}$, $C^2_{e\mu\tau}$, and $\pi_{e\mu\tau}$, respectively by 
\begin{eqnarray}
\tau_{e\mu\tau}&=&\tau_{e(\mu\tau)}-\tau_{e\mu}-\tau_{e\tau}\nonumber \\C^2_{e\mu\tau}&=&C^{2} _{e(\mu\tau)}-C^{2} _{e\mu}-C^{2} _{e\tau}  \\\pi_{e\mu\tau}&=&\frac{1}{3}\left(N^2 _{e(\mu\tau)}+N^2 _{\mu(e\tau)}\right. \nonumber \\ && \hspace{2em} \left. +N^2 _{\tau(e\mu)}  -2N^2 _{e\mu}-2N^2 _{e\tau}-2N^2 _{\mu\tau}\right).\nonumber \end{eqnarray} 
 These quantities represent a collective property of three flavor modes of an electron-neutrino flavor state in three-qubit system that is unchanged by permutations, similar terms for muon and $\tau$ neutrinos can also be defined \cite{Ou_2007}.

The tangle between $e$ and $\mu$ flavor modes  $\tau_{e\mu}$ and between $e$ and $\tau$ flavor modes $\tau_{e\tau}$ is found by calculating reduced density matrix  $\rho^{e\mu}(t)=Tr_{\tau}(\rho^{e\mu\tau}(t))$ and $\rho^{e\tau}(t)=Tr_{\mu}(\rho^{e\mu\tau}(t))$, respectively. For $e\tau$ flavor modes, the eigen values of the product $\rho^{e\tau}(t)\tilde{\rho}^{e\tau}(t)$ are $\lambda_1=\lambda_2=\lambda_3=0$ and $ \lambda_4=4\vert{\tilde{U}_{ee}(t)}\vert^2 \vert{\tilde{U}_{e\tau} (t)}\vert^2$, where $\tilde{\rho}^{e\tau}(t)$ is a \textquotedblleft {spin-flipped} \textquotedblright density matrix $\tilde{\rho}^{e\tau}(t)=(\sigma_y \otimes \sigma_y)\rho^{*e\tau}(t)(\sigma_y \otimes \sigma_y)$. This leads to the tangle for $e\tau$ and similarly, for $e\mu$ flavor modes given by ,
 \begin{eqnarray}
\tau_{e\tau}&=&Tr(\rho^{e\tau}(t)\tilde{\rho}^{e\tau}(t))=4\vert{\tilde{U}_{ee}(t)}\vert^2 \vert{\tilde{U}_{e\tau} (t)}\vert^2,\nonumber \\ 
\tau_{e\mu}&=&Tr(\rho^{e\mu}(t)\tilde{\rho}^{e\mu}(t))=4\vert{\tilde{U}_{ee}(t)}\vert^2 \vert{\tilde{U}_{e\mu} (t)}\vert^2.
\end{eqnarray}
The CKW inequality in terms of tangle is: $\tau_{e\mu}+\tau_{e\tau}=\tau_{e(\mu\tau)}$ and is unchanged by permutation ( i.e, $\tau_{\mu e}+\tau_{\mu\tau}=\tau_{\mu(e\tau)}$, $\tau_{\tau e}+\tau_{\tau\mu}=\tau_{\tau(e\mu)}).$ 

Fig 2. shows the time evolution of the tangle between flavor  mode $e$ and $\mu$ , $e$ and $\tau$ and $e$ and $\mu\tau$ (i.e, $\tau_{e\mu}, \tau_{e\tau}, \tau_{e(\mu\tau)}$), but the residual tangle between flavor mode $e$, $\mu$ and $\tau$ is zero i.e, $\tau_{e\mu\tau}=0$. This means that the flavor neutrino state is in a biseparable state. The result shows that for any values of the tangle satisfying equality $\tau_{e\mu}+\tau_{e\tau}=\tau_{e(\mu\tau)}$, there is a quantum state that is consistent with those values. Similarly $C^2 _{e\mu}+C^2 _{e\tau}=C^2 _{e(\mu\tau)}$ is also unchanged by permutation, therefore residue in concurrence is also zero i.e, $C^2_{e\mu\tau}=0$.
\begin{figure}[ht!] 
        \centering \includegraphics[width=0.7\columnwidth]{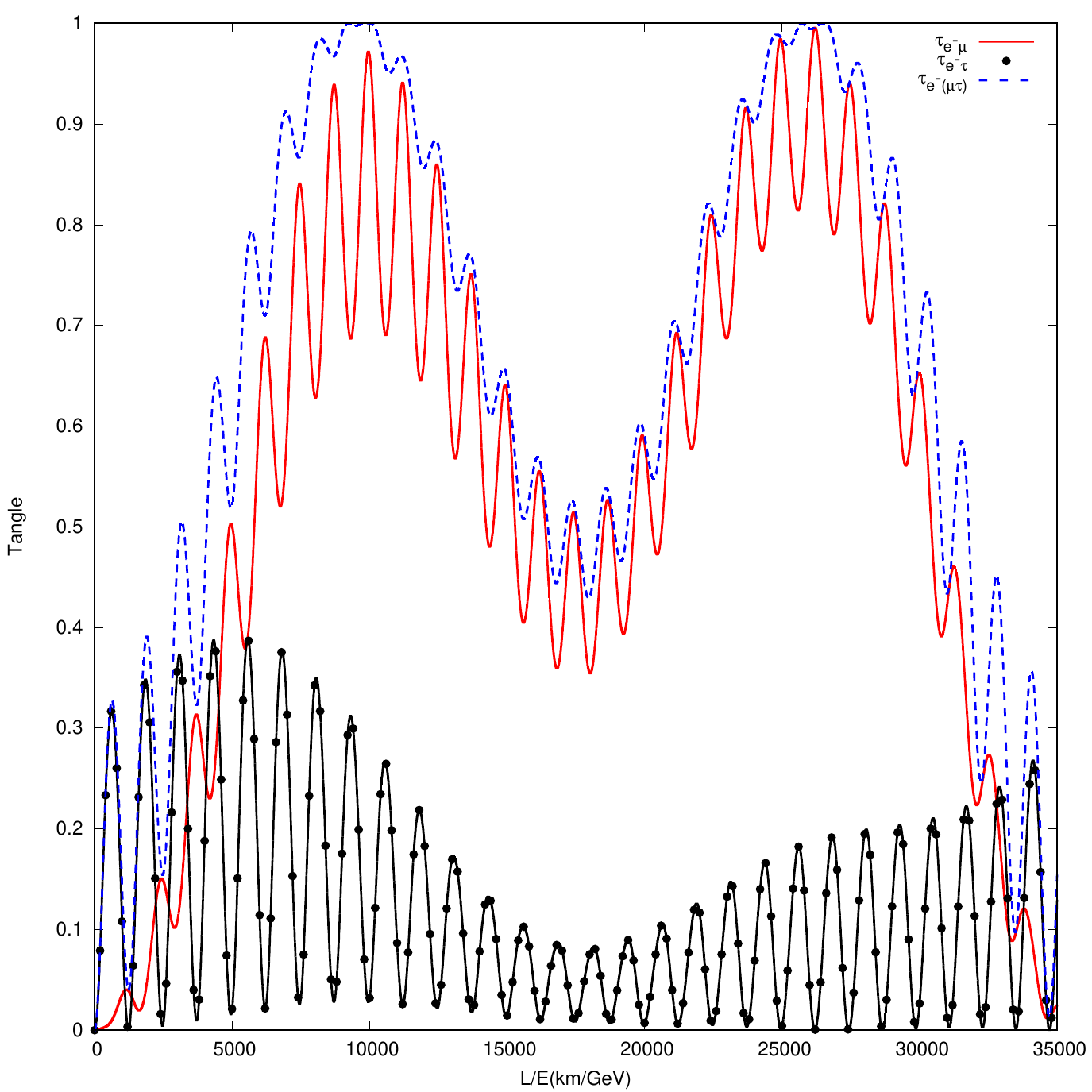}    
        \caption{(Color online) Tangle($\tau$) vs $\frac{L}{E}(\frac{Km}{GeV})$ graph between flavor modes electron, muon and tau neutrinos showing $\tau_{e\mu}+\tau_{e\tau}=\tau_{e(\mu\tau)}$ where representation $\tau_{e\mu}$(Red line), $\tau_{e\tau}$ (Black line), and $\tau_{e(\mu\tau)}$ (Blue line). Parameters $\theta_{ij}$ and $\Delta m^2_{ij}$ are fixed at the experimental values \cite{Esteban:2018azc}.}
\label{fig:2}
\end{figure}

$N_{e\mu}$ and $N_{e\tau}$ are the negtativities of the mixed states $\rho^{e\mu}(t)=Tr_{\tau}(\rho^{e\mu\tau}(t))$ and $\rho^{e\tau}(t)=Tr_{\mu}(\rho^{e\mu\tau}(t))$, respectively \cite{Ou_2007}. We find that the entanglement negativity of the $e$ $\mu$ flavor modes is 
\begin{eqnarray}
\hspace{-1em} N^2_{e\mu}&=&4\vert{\tilde{U}_{ee}(t)}\vert^2\vert{\tilde{U}_{e\mu} (t)}\vert^2  +2\vert{\tilde{U}_{e\tau} (t)}\vert^4 \nonumber \\
&-&2\vert{\tilde{U}_{e\tau} (t)}\vert^2\sqrt{\vert{\tilde{U}_{e\tau} (t)}\vert^4 + 4\vert{\tilde{U}_{ee}(t)}\vert^2\vert{\tilde{U}_{e\mu} (t)}\vert^2} .\end{eqnarray}
For $e$ $\tau$ flavor modes  the negativity is \begin{eqnarray}
\hspace{-1em} N^2_{e\tau}&=&4\vert{\tilde{U}_{ee}(t)}\vert^2\vert{\tilde{U}_{e\tau} (t)}\vert^2 +2\vert{\tilde{U}_{e\mu} (t)}\vert^4 \nonumber \\
&-&2\vert{\tilde{U}_{e\mu} (t)}\vert^2\sqrt{\vert{\tilde{U}_{e\mu} (t)}\vert^4 + 4\vert{\tilde{U}_{ee}(t)}\vert^2\vert{\tilde{U}_{e\tau} (t)}\vert^2}, \end{eqnarray} and also, for the $e$ and $(\mu\tau)$ system we have $N^2 _{e(\mu\tau)}=4\vert{\tilde{U}_{ee}(t)}\vert^2 (\vert{\tilde{U}_{e\mu} (t)}\vert^2 + \vert{\tilde{U}_{e\tau} (t)}\vert^2)$.
The resulting CKW inequality: $N^{2} _{e\mu}+N^{2} _{e\tau}\leq N^{2} _{e(\mu\tau)}$ implies:
\begin{eqnarray}
\hspace{-1em} &&\vert{\tilde{U}_{e\mu} (t)}\vert^4 + \vert{\tilde{U}_{e\tau} (t)}\vert^4 \nonumber \\ && \hspace{1em} < \vert{\tilde{U}_{e\tau} (t)}\vert^2\sqrt{\vert{\tilde{U}_{e\tau} (t)}\vert^4 + 4\vert{\tilde{U}_{ee}(t)}\vert^2\vert{\tilde{U}_{e\mu} (t)}\vert^2}  \nonumber \\ && \hspace{2em}
+\vert{\tilde{U}_{e\mu} (t)}\vert^2\sqrt{\vert{\tilde{U}_{e\mu} (t)}\vert^4 + 4\vert{\tilde{U}_{ee}(t)}\vert^2\vert{\tilde{U}_{e\tau} (t)}\vert^2}.
 \end{eqnarray}
 Fig 3. shows that the time evolution of  the sum of the entanglement negativity between flavor mode $e$ and $\mu$ and between $e$ and $\tau$ is less than entanglement negativity between flavor mode $e$ and $\mu\tau$ i.e, $N^2_{e\mu}+N^2_{e\tau}< N^2_{e(\mu\tau)}$. With this result we can say that the CKW inequality in terms of negativity is strict (because $U_{ee}(t)\neq 0$, $U_{e\mu}(t)\neq 0$, $U_{e\tau}\neq0$) and that, the inequality in terms of concurrence and tangle between different flavor modes of neutrino is characteristic of a general class of W-states.

 \begin{figure}[ht!] 
        \centering \includegraphics[width=0.7\columnwidth]{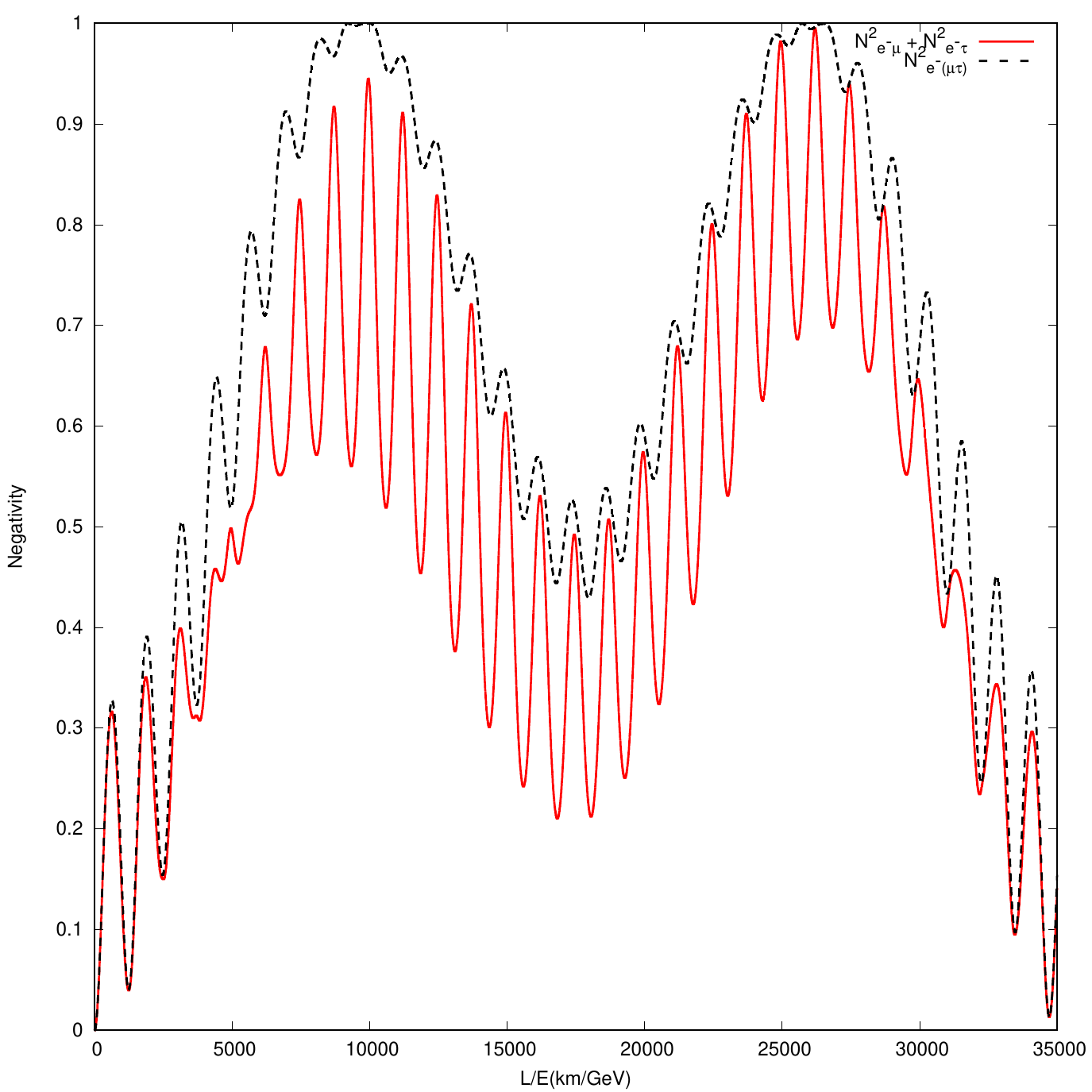}    
        \caption{(Color online) Negativity ($N^2_{e\mu} + N^2_{e\tau}$) (Red line) and $N^2_{e(\mu\tau)}$ (Black line) vs $\frac{L}{E}(\frac{Km}{GeV})$ graph between flavor modes electron, muon, and tau neutrinos satisfying: $N^2_{e\mu}+N^2_{e\tau}< N^2_{e(\mu\tau)}$. Parameters $\theta_{ij}$ and $\Delta m^2_{ij}$ are fixed at the experimental values \cite{Esteban:2018azc}.}
\label{fig:3}
\end{figure}

In order to understand the tightness of the monogamy inequality in terms of negativity, the three-$\pi$ analogous to three tangle ($\tau_{e\mu\tau}$) is studied in the context of three flavor neutrino oscillations. Three-$\pi$  is a natural entanglement measure, which satisfy three necessary conditions:
\begin{itemize}\item It should be local unitary (LU) invariant 
\item It has zero value for product pure states, and 
\item It has value greater than zero for genuine tripartite entanglement \cite{Vedral:1997qn, Ou_2007}.
\end{itemize}

Thus, this non-zero value of residual entanglement three-$\pi$ will be the measure of genuine three particle entanglement. 

For electron-neutrino flavor (and analogously for a muon or tau neutrino system), it can be defined as:
$\pi_{e\mu\tau} = \frac{\pi_e + \pi_\mu +\pi_\tau}{3}$, where
$\pi_e =N^2 _{e(\mu\tau)} -N^2 _{e\mu}-N^2 _{e\tau}$,
$\pi_{\mu} 
=N^2 _{\mu(e\tau)} -
N^2 _{\mu e}
-N^2 _{\mu\tau}$,
 and $\pi_{\tau} =N^2 _{\tau(e\mu)} -N^2 _{\tau e}-N^2 _{\tau\mu}$ are the residual entanglement in terms of negativity and the subscript $e$, $\mu$ and $\tau$ in $\pi_e$, $\pi_\mu$, $\pi_\tau$ mean the flavor mode $e$, flavor mode $\mu$, and flavor mode $\tau$ are taken as the focus respectively.
Using the negativity values calculated earlier to get  $\pi_e$, $\pi_\mu$, and $\pi_\tau$ we find $\pi_e\neq\pi_\mu\neq\pi_\tau$. 
We can see from Fig.4 that unlike tangle and concurrence, the residual entanglement have the different maxima ($\pi_e$, $\pi_\mu$ and $\pi_\tau$) at scale of distance per energy $\frac{L}{E}>0$, and  $\pi_e\neq\pi_\mu\neq\pi_\tau$. This gives clear indication that the residual entanglement $\pi_e$, $\pi_\mu$ and $\pi_\tau$ are quantified  but it can not be the measure of genuine tripartite entanglement as the measures are not invariant under permutations. As the measure of genuine tripartite entanglement in three flavor neutrino oscillations, we define  $\pi_{e\mu\tau}$ as the average of $\pi_e$, $\pi_\mu$, and $\pi_\tau$, such that $\pi_{e\mu\tau}=\frac{1}{3}(N^2 _{e(\mu\tau)}+N^2 _{\mu(e\tau)}+N^2 _{\tau(e\mu)}-2N^2 _{e\mu}-2N^2 _{e\tau}-2N^2 _{\mu\tau})$. 
$\pi_{e\mu\tau}$ is now invariant under permutations of flavor mode in an electron- neutrino flavor state. Thus, 
\begin{eqnarray}
\pi_{e\mu\tau}&=&
\frac{4}{3}[\vert{\tilde{U}_{ee} (t)}\vert^2\sqrt{\vert{\tilde{U}_{ee} (t)}\vert^4 + 4\vert{\tilde{U}_{ee}(t)}\vert^2\vert{\tilde{U}_{e\tau} (t)}\vert^2} \nonumber \\
&+& \vert{\tilde{U}_{e\mu} (t)}\vert^2\sqrt{\vert{\tilde{U}_{e\mu} (t)}\vert^4+4\vert{\tilde{U}_{ee}(t)}\vert^2\vert{\tilde{U}_{e\tau} (t)}\vert^2} \nonumber \\ &+& \vert{\tilde{U}_{e\tau} (t)}\vert^2 \sqrt{\vert{\tilde{U}_{e\tau} (t)}\vert^4 + 4\vert{\tilde{U}_{ee}(t)}\vert^2\vert{\tilde{U}_{e\mu} (t)}\vert^2} \nonumber \\
&-&\vert{\tilde{U}_{ee} (t)}\vert^4 -\vert{\tilde{U}_{e\mu} (t)}\vert^4
-\vert{\tilde{U}_{e\tau} (t)}\vert^4]
>0.
\end{eqnarray}

\begin{figure}[ht!] 
        \centering \includegraphics[width=0.7\columnwidth]{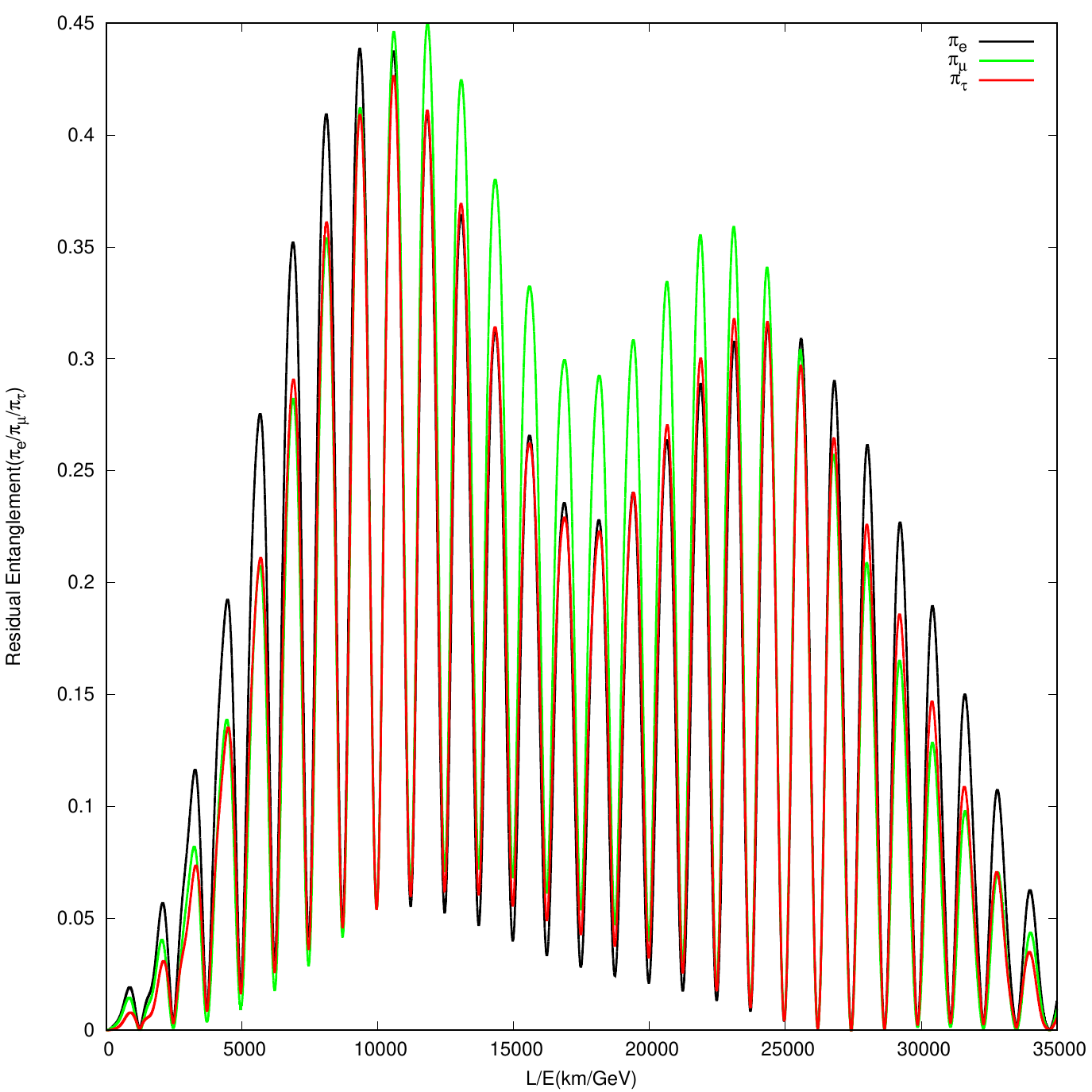}    
        \caption{(Color online) Residual entanglement $\pi_{e}$ (Black line), $\pi_{\mu}$ (Green line), $\pi_{\tau}$ (Red line) vs  $\frac{L}{E}(\frac{Km}{GeV})$ graph between flavor modes of electron, muon and tau neutrinos satisfying: $\pi_e\neq \pi_\mu \neq \pi_\tau$. Parameters $\theta_{ij}$ and $\Delta m^2_{ij}$ are fixed at the experimental values \cite{Esteban:2018azc}.} 
\label{fig:4}
\end{figure}

From Fig.5, we note that for $\frac{L}{E}>0$, entanglement among three-flavor modes occurs i.e, $\pi_{e\mu\tau}>0$, and exhibits a typical oscillatory behavior. At largest mixing i.e, when transition probabilities are $P_{\nu_{e\rightarrow e}}=0.39602$, $P_{\nu_{e\rightarrow \mu}}=0.435899$, and  $P_{\nu_{e\rightarrow\tau}}=0.168081$, we find that $\pi_{e\mu\tau}$ reaches the maximum value 0.436629.

\begin{figure}[ht!] 
        \centering \includegraphics[width=0.7\columnwidth]{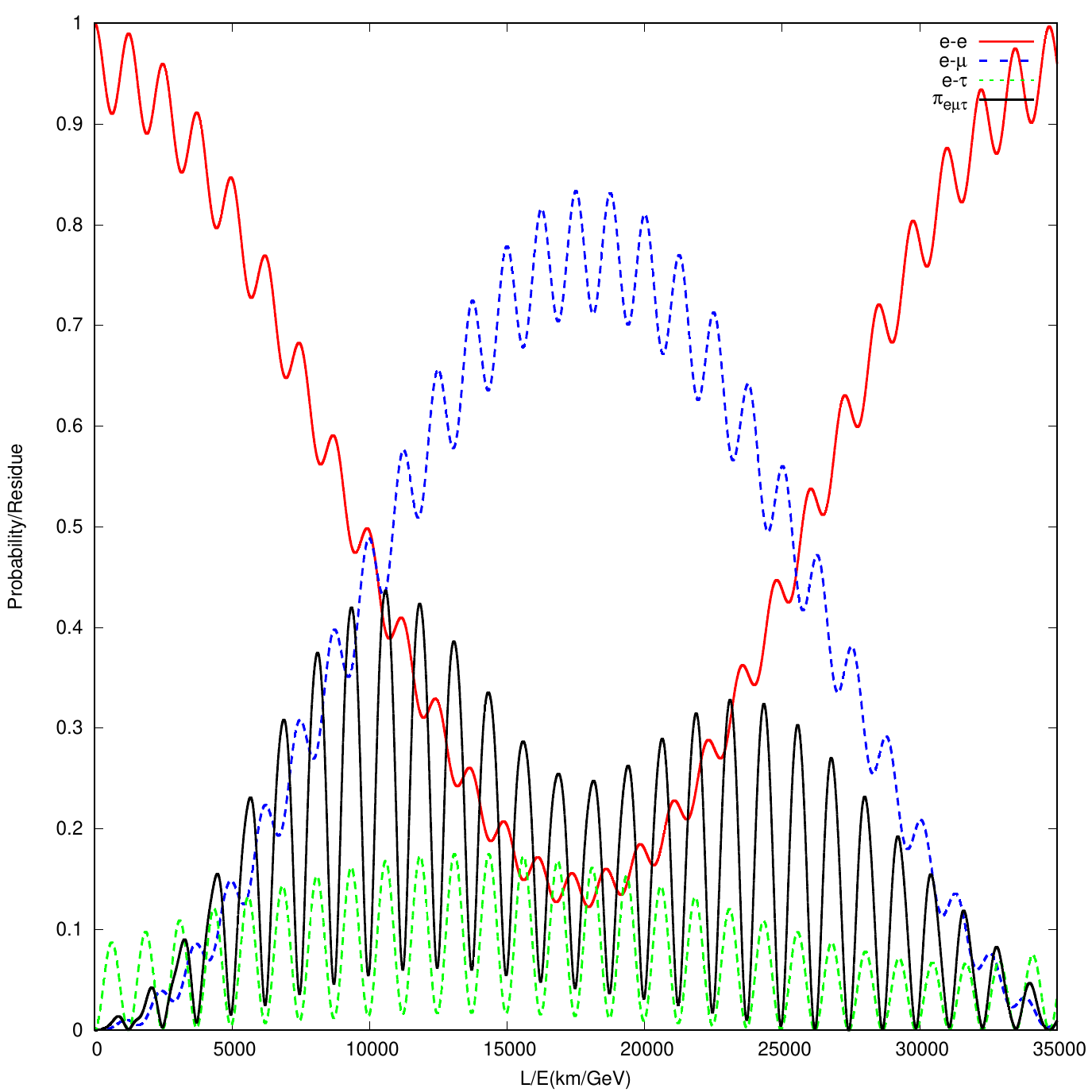}    
        \caption{(Color online) Residual entanglement $\pi_{e\mu\tau}$ (Black line) vs $\frac{L}{E}(\frac{Km}{GeV})$ graph between the flavor modes electron, muon and tau neutrinos. Parameters $\theta_{ij}$ and $\Delta m^2_{ij}$ are fixed at the experimental values\cite{Esteban:2018azc}. The transition probabilities $P_{\nu_{e\rightarrow e}}$ (Red line), $P_{\nu_{e\rightarrow \mu}}$ (Blue line) and  $P_{\nu_{e\rightarrow\tau}}$ (Green line) are reported as well for comparison.}
\label{fig:5}
\end{figure}
 
 So far, we have considered the time evolution of entanglement characteristics of an electron neutrino state, which are relevant for reactor experiments. For completeness, we give the relevant entanglement measures for a muon neutrino state relevant to accelerator experiments.
  \begin{equation}
 \ket{\nu_{\mu}(t)}=\tilde{U}_{\mu e}(t)\ket{100}_e+\tilde{U}_{\mu\mu }(t)\ket{010}_\mu+\tilde{U}_{\mu\tau}\ket{001}_\tau
  \end{equation}
  where, $\vert{\tilde{U}_{\mu e}(t)}\vert^2+\vert{\tilde{U}_{\mu\mu}(t)}\vert^2+\vert{\tilde{U}_{\mu\tau}(t)}\vert^2=1$. The relevant density matrix is $\rho^{\mu e \tau}(t)$ for the initial muon-flavor neutrino state the CKW inequality in terms of tangle and concurrence becomes equal, consequently the residual tangle and concurrence vanishes, i.e $\tau_{\mu e \tau}=C^2_{\mu e \tau}=0$. Whereas, the CKW inequality in terms negativity is strict i.e, $N^2_{\mu e}+ N^2_{\mu\tau} < N^2_{\mu (e\tau)}$.
  \begin{figure}[ht!] 
        \centering \includegraphics[width=0.7\columnwidth]{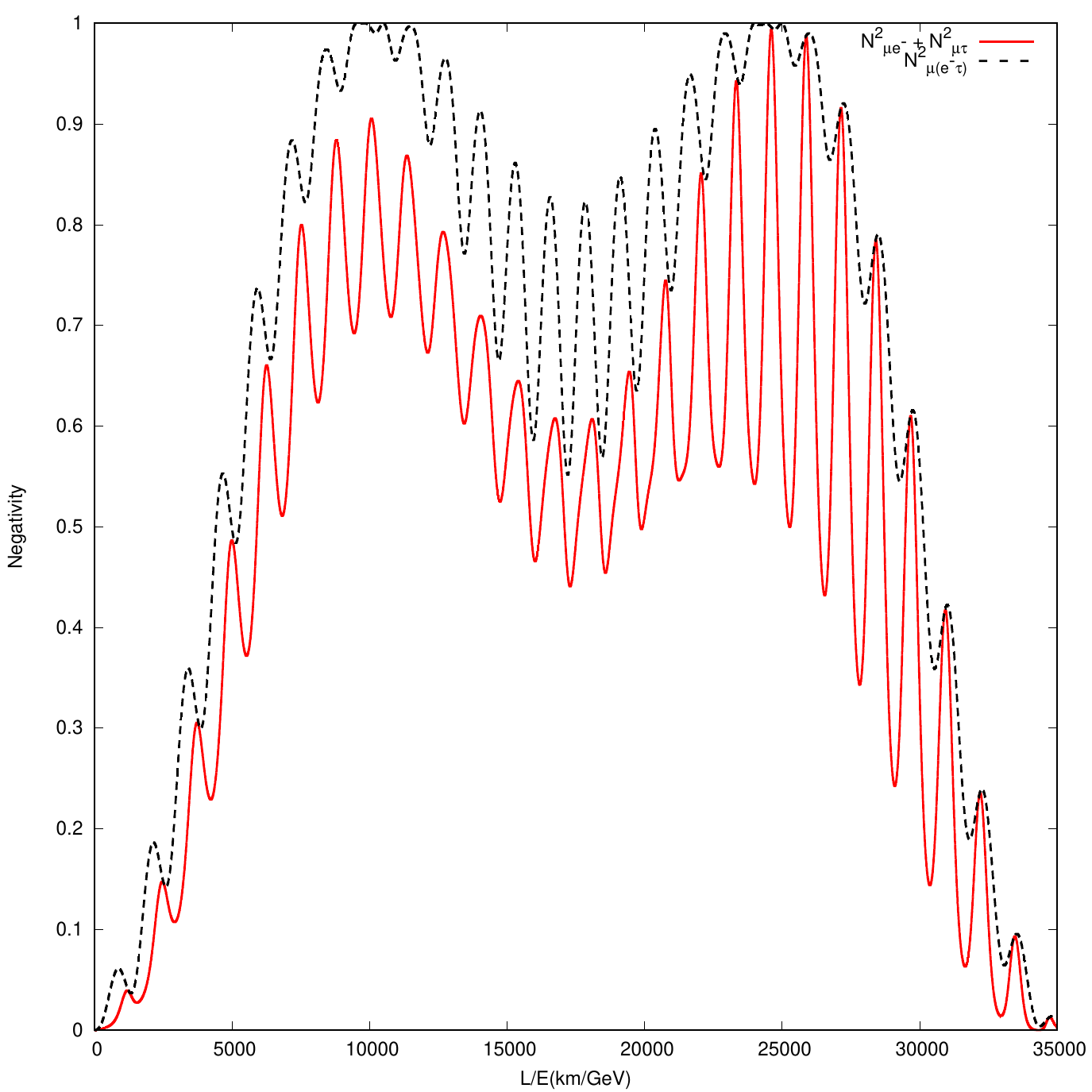}    
        \caption{ (Color online) Negativity ($N^2_{\mu e} + N^2_{\mu\tau}$) (Red line) and $N^2_{\mu(e\tau)}$ (Black line) vs $\frac{L}{E}(\frac{Km}{GeV})$ graph between flavor modes electron, muon, and tau neutrinos satisfying: $N^2_{\mu e}+N^2_{\mu\tau}< N^2_{\mu(e \tau)}$. Parameters $\theta_{ij}$ and $\Delta m^2_{ij}$ are fixed at the experimental values \cite{Esteban:2018azc}.}
\label{fig:6}
\end{figure}
 In Fig.6, it is seen that the sum of the negativity between flavor modes $\mu$ and $e$, and between $\mu$ and $\tau$ is less than the negativity between flavor modes $\mu$ and $e \tau$. The measure of tripartite entanglement is 
 \begin{eqnarray}
 \pi_{\mu e \tau}&=&
\frac{4}{3}[\vert{\tilde{U}_{\mu e} (t)}\vert^2\sqrt{\vert{\tilde{U}_{\mu e} (t)}\vert^4 + 4\vert{\tilde{U}_{\mu e}(t)}\vert^2\vert{\tilde{U}_{\mu \tau} (t)}\vert^2} \nonumber \\
&+& \vert{\tilde{U}_{\mu\mu} (t)}\vert^2\sqrt{\vert{\tilde{U}_{\mu\mu} (t)}\vert^4+4\vert{\tilde{U}_{\mu e}(t)}\vert^2\vert{\tilde{U}_{\mu \tau} (t)}\vert^2} \nonumber \\ &+& \vert{\tilde{U}_{\mu\tau} (t)}\vert^2 \sqrt{\vert{\tilde{U}_{\mu\tau} (t)}\vert^4 + 4\vert{\tilde{U}_{\mu e}(t)}\vert^2\vert{\tilde{U}_{\mu\mu} (t)}\vert^2}\nonumber \\
&-&\vert{\tilde{U}_{\mu e} (t)}\vert^4 -\vert{\tilde{U}_{\mu \mu} (t)}\vert^4
-\vert{\tilde{U}_{\mu\tau} (t)}\vert^4]
>0.
\end{eqnarray}

\begin{figure}[ht!] 
        \centering \includegraphics[width=0.7\columnwidth]{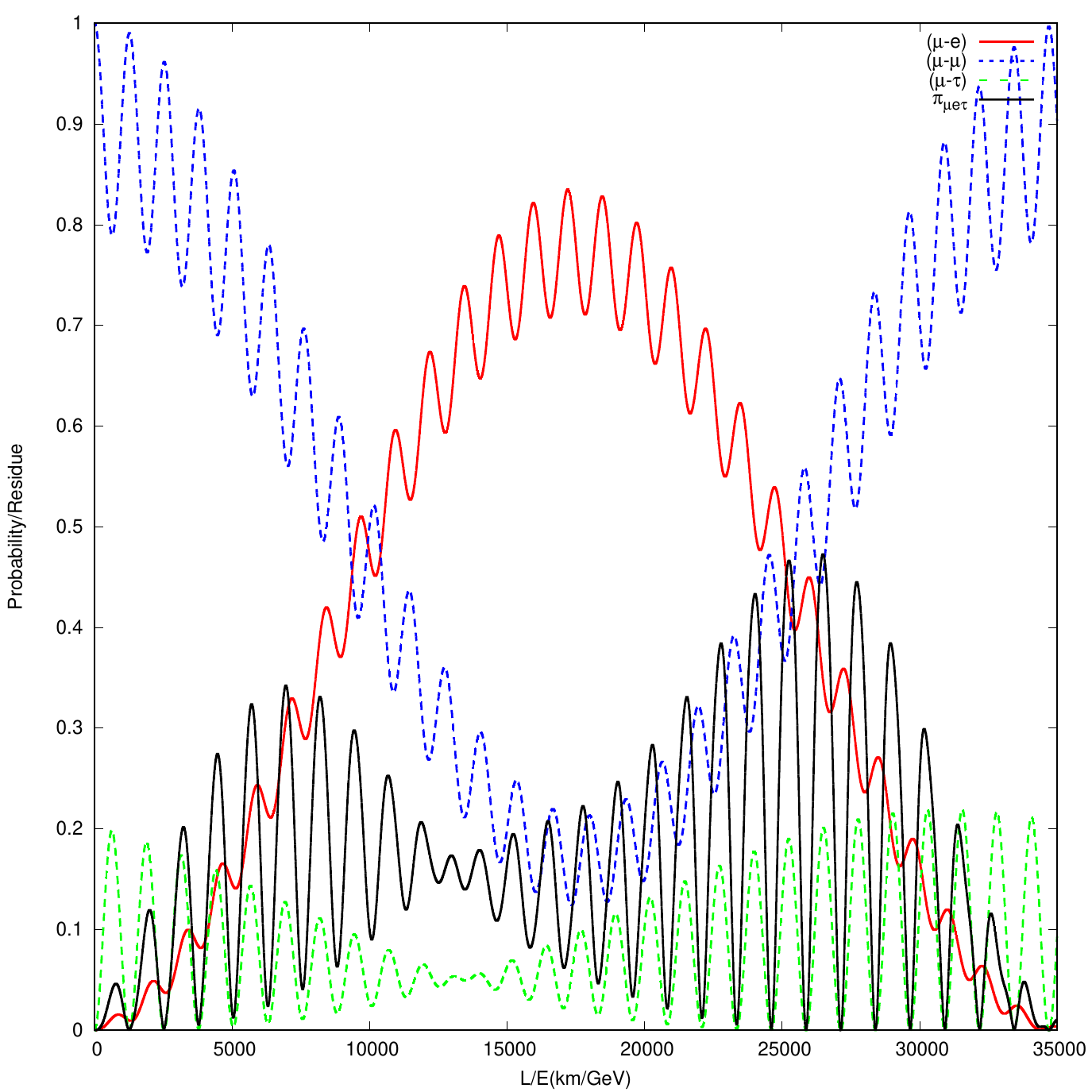}    
        \caption{(Color online) Residual entanglement $\pi_{\mu e \tau}$ (Black line) vs $\frac{L}{E}(\frac{Km}{GeV})$ graph between the flavor modes electron, muon and tau neutrinos. Parameters $\theta_{ij}$ and $\Delta m^2_{ij}$ are fixed at the experimental values \cite{Esteban:2018azc}. The transition probabilities $P_{\nu_{\mu\rightarrow e}}$ (Red line), $P_{\nu_{\mu\rightarrow \mu}}$ (Blue line)  and  $P_{\nu_{\mu\rightarrow\tau}}$ (Green line) are reported as well for comparison.}
\label{fig:7}
\end{figure}

From Fig.7, we observe that at $\frac{L}{E}=0$, $\pi_{\mu e \tau}=0$, which mean initial muon-neutrino flavor state is a biseparable state. At $\frac{L}{E}>0$, entanglement among three-flavor modes occur i.e, $\pi_{\mu e \tau}>0$, exhibits a typical oscillatory behaviour. Therefore the entanglement between flavor modes is now neither separable nor biseparable. At largest mixing $\pi_{\mu e \tau}$ reaches the maximum value 0.472629 indicating genuine tripartite entanglement.

 The fact that the three neutrino state exhibits genuine bipartite entanglement leads us to the identification of this state with the W-state of quantum optics.
 The prototype W-state is  $\ket{W_1}=\frac{1}{\sqrt{3}}(\ket{100}+\ket{010}+\ket{001})$, and the maximum value of three-$\pi$ value goes to $\pi_{ABC}(W_1)=\frac{4}{9}(\sqrt{5}-1)=0.549363$ \cite{ Ou_2007}. For the class of W-state as an entanglement resource, $\ket{W_2}=\frac{1}{2}(\ket{100}+\ket{010}+\sqrt{2}\ket{001})$, with $\pi_{ABC}({W_2})=0.47140$ \cite{Pati:2006}. Comparing the value of $\pi_{e\mu\tau}$ and $\pi_{\mu e \tau}$ with the three-$\pi$ value of different class of W-states, we get 
 \begin{equation}
  \pi_{e\mu\tau}(\nu_e)<\pi_{ABC}(W_2)<\pi_{\mu e \tau}(\nu_\mu)<\pi_{ABC}(W_1).
  \end{equation} 
 Hence, satisfying CKW inequality and with all properties of W-state 
\begin{eqnarray}
&& \pi_{e\mu\tau}>\tau_{e\mu\tau}=C^{2}_{e\mu\tau}=0 \nonumber\\ \mbox{or,} \hspace{1em} && \pi_{\mu e \tau}>\tau_{\mu e \tau}=C^{2}_{\mu e \tau}=0, 
 \end{eqnarray} 
 implies that the form of mode (flavor) entangled neutrino state Eq.(5) has the general properties of tri-partite entangled W-state. 
 
\section{CONCLUSION}
In conclusion, we have explored various entanglement measures of neutrino flavor oscillations in bipartite and tripartite quantum states and compared our results with the two qubit and three qubit classes of W-state in quantum optics. In bipartite quantum system, all quantum correlations like tangle, concurrence, negativity coincides with the linear entropy (a lower approximation to the von Neumann entropy). It reveals that the $\ket{\nu_{e}(t)}$ is a bipartite entangled pure state. In three qubit case, three flavor neutrino oscillation satisfies CKW inequality and exhibits the property of the class of W-states. Consequently, $\pi_{e\mu\tau}>0$ or $\pi_{\mu e \tau}>0$ implies a generalized form of genuine tripartite entanglement in three flavor neutrino oscillations. Using the fact that the neutrino mixing in the two mode case is akin to entanglement via mode swapping that takes place due to a BS, we conjecture that a quantum optical sytem using a collection of beam spliters to mimic the generalized W state akin to the three neutrino state can be constructed. This will be the subject of further investigation. 

To investigate the possibility of an experimental signal of neutrino entanglement, the Daya-Bay experiment has analysed  the wave-packet model of neutrino oscillations. A many particle system can be considered in highly entangled wave packet state. The coherent evolution of the electron neutrino state and subsequent decoherence, is the subject of a recent  experimental paper \cite{An:2016pvi}. Quantum coherence in experimentally observed neutrino oscillations, using the tools of quantum resource theory, have produced results for the longest distance over which quantumness has been experimentally determined for quantum particles other than photon \cite{Song:2018bma}. In fact, flavor oscillations and entanglement in the "strange" $K^0$-$\bar{K^0}$ system had been already investigated in the last century \cite{Bertlmann:2004yg}. It is of interest for future experiments to give justification for three way entanglement in neutrino oscillations and see how to explore it further in  quantum information systems. Since quantum optical systems can be manipulated, unlike neutrino oscillations experiments, the work done by us is of interest to explore the characteristics of neutrino oscillation quantum entanglement further. Work on simulation of such systems on a quantum computer is in progress.
\section{Acknowledgments}
We would like to thank Prof. C. Mukku for his insightful inputs. AKJ and SM acknowledge fruitful discussions with Prof. S. Umashankar of IIT, Mumbai. SM acknowledge partial financial support of the UGC-Networking Resource Centre, School of Physics, University of Hyderabad, India. We would also like to thank Akshay Chatla for his valuable suggestions.

\end{document}